\documentclass[iop,twocolappendix,numberedappendix]{emulateapj}
\usepackage{epsfig}

\def\gtsima{$\; \buildrel > \over \sim \;$}
\def\ltsima{$\; \buildrel < \over \sim \;$}
\def\gtrsim{\lower.5ex\hbox{\gtsima}}
\def\lesssim{\lower.5ex\hbox{\ltsima}}

\shorttitle{Planets and protoplanets in the Galactic center}
\shortauthors{Mapelli \&{} Ripamonti}

\begin{document}

\title{Signatures of planets and protoplanets in the Galactic center: a clue to understand the G2 cloud?}
\author{Michela Mapelli\altaffilmark{1} \&{} Emanuele Ripamonti\altaffilmark{2,1}}
\altaffiltext{1}{INAF-Osservatorio Astronomico di Padova, Vicolo dell'Osservatorio 5, I--35122, Padova, Italy; {\tt michela.mapelli@oapd.inaf.it}}
\altaffiltext{2}{Dipartimento di Fisica e Astronomia Galileo Galilei, University of Padova, Vicolo dell'Osservatorio 3,  I--35122, Padova, Italy}

\begin{abstract}
Several hundred young stars lie in the innermost parsec of our Galaxy. The super-massive black hole (SMBH) might capture planets orbiting these stars, and bring them onto nearly radial orbits. The same fate might occur to planetary embryos (PEs), i.e. protoplanets born from gravitational instabilities in protoplanetary disks. In this paper, we investigate the emission properties of rogue planets and PEs in the Galactic center.  In particular, we study the effects of photoevaporation, caused by the ultraviolet background. Rogue planets can hardly be detected by current or forthcoming facilities, unless they are tidally disrupted and accrete onto the SMBH. In contrast, photoevaporation of PEs (especially if the PE is being tidally stripped) might lead to a recombination rate as high as $\approx{}10^{45}$ s$^{-1}$, corresponding to a Brackett-$\gamma{}$ luminosity $L_{\rm Br-\gamma{}}\approx{}10^{31}$ erg s$^{-1}$, very similar to the observed luminosity of the dusty object G2. We critically discuss the possibility that G2 is a rogue PE, and the major uncertainties of this model.
\end{abstract}


%

\section{Introduction}
The Galactic center (GC) is one of the most studied and yet enigmatic places in our Universe. It is exceptionally crowded: it hosts a supermassive black hole (SMBH, \citealt{Gillessen2009}), a dense star cluster of old stars (\citealt{Schoedel2007}), several thousand solar masses of molecular, atomic and ionized gas (\citealt{baobab12,Jackson93,Scoville03}), and a few hundred young stars (\citealt{Genzel03,Paumard06,Bartko09,Lu09,Lu13}). A relevant fraction ($\sim{}20$\%) of the young massive stars lie in the so called clockwise disk (\citealt{Paumard06,Bartko09,Lu09,Lu13}), a thin disk with inner radius $\sim{}0.04$ pc and outer radius $\sim{}0.13$ pc (\citealt{Yelda14}). A small group of B-type stars ($\sim{}20$ objects, forming the so called S-cluster) orbit around the SMBH with semi-major axes $<0.04$ pc, with isotropically oriented orbital planes, and with high eccentricities (approximately following a thermal distribution, \citealt{Gillessen09b}). The formation of the young stars in the clockwise disk and in the S-cluster has been a puzzle for a long time, because the tidal shear from the SMBH disrupts molecular clouds, preventing star formation in normal conditions. 

A faint dusty object (named G2, \citealt{Gillessen12}) has been observed on a very eccentric orbit around the SMBH, with a periapse distance of only $\sim{}200$ AU (\citealt{Witzel14, Pfuhl14}). G2 has transited at periapse in Spring 2014, avoiding complete tidal disruption. The nature of G2 is still enigmatic:  a pure gas cloud (\citealt{Gillessen12,Schartmann12,Burkert12,Shcherbakov14,DeColle14,McCourt14}), a dust-enshrouded low-mass star (\citealt{Burkert13,Ballone13,Witzel14}), a low-mass star with a protoplanetary disk  (\citealt{Murray-ClayLoeb13}), a star disrupted by a stellar-mass black hole (\citealt{Miralda-Escude}), a star that underwent partial tidal disruption by the SMBH (\citealt{Guillochon14}), a merger between two stars (\citealt{Prodan15}), and a nova outburst (\citealt{Meyer12}) have been proposed to explain its properties (see \citealt{MapelliGualandris15} for a recent review).

Planets have not been detected in the GC so far, but the destiny of planets, asteroids and planetesimals has been investigated by several authors.  Collisions of planets or asteroids have been proposed to lead to the formation of a dusty torus around SMBHs (\citealt{Nayakshin12}). The tidal disruption of planetesimals by the SMBH has been invoked as mechanism to explain the daily infrared flares of SgrA$^{\ast{}}$ (\citealt{Cadez08, Kostic09, Zubovas12, Hamers14}).  Tidal disruptions of planets are expected to be much less frequent, but more dramatic events, and might account for the possible past activity of SgrA$^{\ast{}}$ (\citealt{Revnivtsev04, Terrier10, Ponti10, Zubovas12}). A system composed of a low-mass star and its protoplanetary disk is one of the most viable scenarios to explain the G2 object (\citealt{Murray-ClayLoeb13}).  Recently, radio-continuum observations revealed 44 partially resolved compact sources in the innermost $\sim{}0.1$ pc, interpreted as candidate photoevaporative protoplanetary disks \citep{Yusef15}.

The aim of this paper is to investigate the main possible signatures of planets and planetary embryos (PEs, i.e. dense gas clouds produced by local gravitational instabilities in a protoplanetary disk, \citealt{Kuiper51,Cameron78,Boss97,Durisen07}) in the GC. In Sections~\ref{sec:planet} and \ref{sec:PE}, we describe the mechanisms that might produce rogue planets and PEs in the GC, and we estimate the mass loss that planets and PEs undergo because of photoevaporation by the ultraviolet (UV) background. In Section~\ref{sec:discussion}, we  discuss the observational signatures of planets and PEs, with particular attention for the Br$\gamma{}$ line emission), and we suggest that the G2 object might be associated with a rogue PE. Section~\ref{sec:summary} is a summary of our main results.
\section{Rogue planets}\label{sec:planet}
\subsection{Tidal capture by the SMBH}
\begin{figure}
\center{{
\epsfig{figure=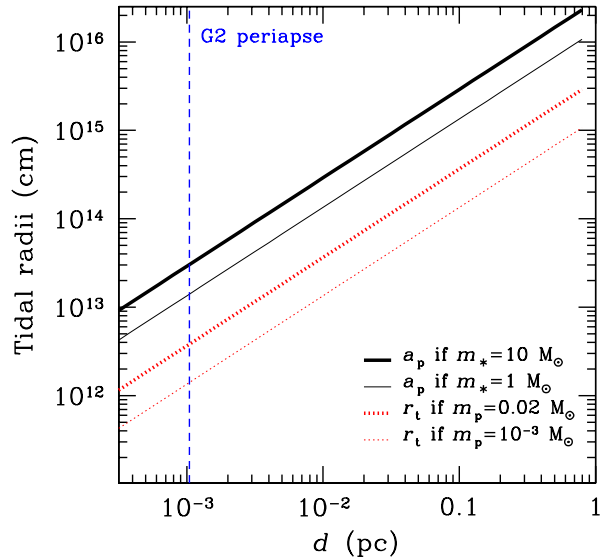,width=8.0cm} 
}}
\caption{\label{fig:fig1}
Tidal radius for splitting a star-planet system (solid black lines) and tidal radius for disruption of the planet (dotted red lines) as  a function of the star-planet distance from the SMBH. Thick and thin solid black lines correspond to a star mass $m_\ast{}=10$ and 1 M$_\odot$, respectively. Thick and thin dotted red lines correspond to a mass $m_{\rm p}=0.02$  M$_\odot$ (i.e. a brown dwarf) and $10^{-3}$ M$_\odot$ (i.e. a Jupiter-like planet), respectively.  The vertical dashed line is the estimated periapse distance of the G2 object. 
}
\end{figure}

The tidal shear of the SMBH can unbind a planet from its star if the initial semi-major axis of the planet orbit is
\begin{eqnarray}\label{eq:ap}
a_{\rm p}\ge 
19\,{}{\rm AU}\left(\frac{d}{0.01\,{}{\rm pc}}\right)\,{}\left(\frac{m_{\ast{}}}{10\,{}{\rm M}_\odot}\right)^{1/3}\,{}\left(\frac{4\times{}10^6\,{}{\rm M}_\odot}{M_{\rm BH}}\right)^{1/3}
\end{eqnarray}
where $d$ is the periapse of the star orbit around the SMBH, $m_{\ast{}}$ is the star mass, and $M_{\rm BH}$ is the SMBH mass. Fig.~\ref{fig:fig1} shows $a_{\rm p}$ as a function of $d$, for $m_\ast{}=1$ and 10 M$_\odot{}$.

 One of the two members of the split binary (generally the most massive one) receives a kick that makes it more bound to the SMBH, while the other member (generally the less massive one) becomes less bound. The less bound object might be ejected, while the remaining one is captured  by the SMBH (e.g. \citealt{Ginsburg12}). On the other hand, the typical variation $\delta{}_v$ of the velocity of the planet is (\citealt{Pfahl05})
\begin{eqnarray}
\delta{}_v\sim{}\sqrt{2}\,{}\left(\frac{G\,{}m_\ast{}}{a_{\rm p}}\right)^{1/2}\,{}\left(\frac{M_{\rm BH}}{m_\ast{}}\right)^{1/6}\nonumber\\\sim{}170\, {\rm km \,{}s}^{-1} \,{}\left(\frac{m_\ast{}}{1\,{}{\rm M}_\odot{}}\right)^{1/2}\,{}\left(\frac{10\,{}{\rm AU}}{a_{\rm p}}\right)^{1/2}\,{}\left(\frac{M_{\rm BH}/m_\ast{}}{4\times{}10^{6}}\right)^{1/6},
\end{eqnarray}

since $\delta{}_v$ is lower than the Keplerian velocity around the SMBH at $d=0.01$ pc ($\sim{}1300$ km s$^{-1}$), it is plausible that both the planet and the star remain bound to the SMBH. Dynamical simulations are necessary to quantify how many planets will be ejected and how many will be captured by the SMBH.  If the planet is captured, its semi-major axis $a_{\rm cap}$ and eccentricity $e_{\rm cap}$ would then be (\citealt{Hills91, Perets09})
\begin{eqnarray}\label{eq:cap}
a_{\rm cap}\simeq{}1.30\,{}{\rm pc}\,{}\left(\frac{a_{\rm p}}{19\,{}{\rm AU}}\right)\,{}\left(\frac{M_{\rm BH}}{4\times{}10^6\,{}{\rm M}_\odot}\right)^{2/3}\,{}\left(\frac{1\,{}{\rm M}_\odot}{m_\ast{}}\right)^{2/3},\\
e_{\rm cap}=1-\frac{d}{a_{\rm cap}}\sim{}0.99.\quad{}
\end{eqnarray}

The tidal split of planets from their stars can be substantially enhanced by planet-planet scatterings, which were shown to be able to scatter Jupiter-mass planets on orbits with semi-major axis$>50$ AU (\citealt{Chatterjee08,Marois08,Chatterjee11}). The planet could even be ejected from its initial system by planet-planet scattering: \citet{Veras09}  estimate that $\sim{}40$\% of planets in a multiple-planet system are ejected by planet-planet scatterings in $2\times{}10^{8}$ Myr. The ejection of Jupiter-like planets from their planet systems is supported by the observation of `freely floating' giant planets (\citealt{Sumi11}).

It is even possible that starless planets form directly from gravitational instabilities in accretion disks around SMBHs (\citealt{Shlosman89, Nayakshin06}). Similarly, starless planets could have formed in the same star formation episode that gave birth to the clockwise disk: according to one of the most popular scenarios, a molecular cloud disrupted by the SMBH might have settled into a parsec-scale dense gaseous disk. Gravitational instabilities in the gaseous disk might have led to the formation of the young massive stars that lie in the clockwise disk (e.g. \citealt{Paczynski78,Kolykhalov80,Shlosman87, Collin99,Gammie01, Goodman03, Tan05, Nayakshin07,Bonnell08,Mapelli08,Hobbs09,Alig11,Mapelli12,Mapelli13, Lucas13}),  and also to the formation of starless giant planets (\citealt{Shlosman89}). 
In such case, the initial orbit of the planets would be inside the clockwise disk, and then  gravitational interactions with stars or other planets might plunge the planets on a more radial orbit. For example, \citet{Murray-ClayLoeb13} predict that $\sim{}1/100$ of the entire population of low-mass objects in the clockwise disk could have been delivered onto highly eccentric orbits by two-body interactions with massive stars.

What happens to a planet that is captured by the SMBH on a very eccentric orbit? Tidal disruption is unlikely, as the planet should have a radius $r_{\rm p}$ larger than
\begin{equation}\label{eq:rp}
r_{\rm t}=1.3\times{}10^{13}{\rm cm}\,{}\frac{d}{0.01\,{}{\rm pc}}\,{}\left(\frac{4\times{}10^6\,{}{\rm M}_\odot}{M_{\rm BH}}\right)^{1/3}\,{}\left(\frac{m_{\rm p}}{10^{-3}\,{}{\rm M}_\odot}\right)^{1/3},
\end{equation}
i.e. $\sim{}2000\,{}(d/0.01\,{}{\rm pc})$ times the radius of Jupiter. Fig.~\ref{fig:fig1} shows the behavior of $r_{\rm t}$ as a function of $d$, for a planet mass $m_{\rm p}=10^{-3}$ M$_\odot$ and for a brown dwarf of mass $m_{\rm p}=2\times{}10^{-2}$ M$_\odot$.

\subsection{Photoevaporation}
The planet will suffer continuous atmosphere evaporation by the UV field of the young stars in the central parsec. 
According to \citet{Murray-Clay09} (using the simplified eq. 19), the mass loss rate by atmosphere evaporation is
\begin{equation}
\dot{M}_{\rm MC} \simeq {{\epsilon \pi{}\,{} r_{\rm p}^2\,{} L_{\rm UV} / (4 \,{}\pi{}\,{} D^2)} \over
{G\,{}m_{\rm p} / r_{\rm p}}},
\label{murrayclay19}
\end{equation}
where $r_{\rm p}$ is the planet radius, $L_{\rm UV}\sim{}10^{40}$ erg s$^{-1}$ is the ionizing luminosity of massive stars in the GC, $\epsilon$ is the fraction of the UV luminosity that goes into heat ($\simeq{} \tilde{\phi}/\phi_0-1$, where $\tilde{\phi}$ is the average energy of ionizing photons, and $\phi_0\simeq{}13.6\,{\rm eV}$ is the ionization energy for atomic H), and $D$ is the distance of the massive stars from the rogue planet. We assume that $D\simeq 0.1\, {\rm pc}$, since this is the outer rim of the clockwise disk in the GC (\citealt{Yelda14}). 

\begin{figure}
\center{{
\epsfig{figure=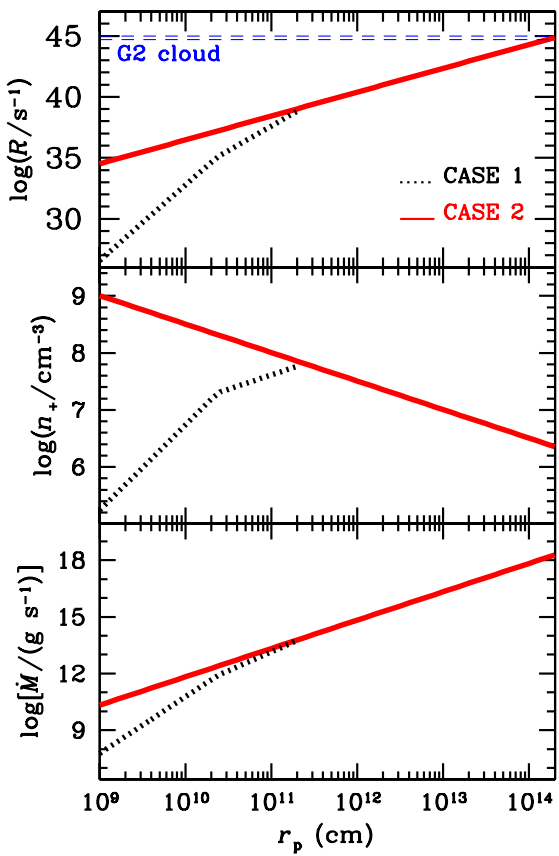,width=8.0cm} 
}}
\caption{\label{fig:fig2}
   From top to bottom:  recombination rate associated with photoevaporation ($R$), evaporating gas density ($n_+$), and mass loss rate by photoevaporation ($\dot{M}$) as a function of the planet (or PE) radius. Black dotted line: CASE~1 (photoevaporating planet, Table~1).  CASE~1 is valid only for $r_{\rm p}\leq{}2\times{}10^{11}$ cm (i.e. for $\tau<1$, see Section~\ref{sec:PEphot} and eq.~\ref{eq:optdepth} for details). Red solid line: CASE~2 (photoevaporating cloud, Table~1).
 The two blue dashed lines in the top panel are the minimum and maximum value of the recombination rate measured for the G2 cloud in the Br$\gamma{}$ line since 2004 (\citealt{Pfuhl14}). 
}
\end{figure}

 For large values of $r_{\rm p}$, eq. (\ref{murrayclay19}) implies that the number of ionizations is larger than the number of ionizing photons reaching the planet surface,
\begin{equation}
Q_{\rm p} = {{\pi{}\,{} r_{\rm p}^2\,{} L_{\rm UV} / (4\,{} \pi{}\,{} D^2)} \over
{\phi_0 / (1-\epsilon)}}.
\label{eq:Qplanet}
\end{equation}
Therefore, we employ an evaporation rate
\begin{eqnarray}\label{eq:mdot}
\dot{M} = \min(\dot{M}_{\rm MC},\,{}m_{\rm prot}\,{}Q_{\rm p}) = \nonumber\\
  = {{r_{\rm p}^2\,{} L_{\rm UV}} \over {4\,{}D^2}} {m_{\rm prot} \over {\phi_0}}
  \min \left[{\epsilon {{\phi_0/m_{\rm prot}} \over {(G\,{}m_{\rm p}) / r_{\rm p}}},
        (1-\epsilon{})}\right] \simeq \nonumber\\
  2.0\times10^{11}\,{\rm g\, s^{-1}}
  \left({r_{\rm p} \over {10^{10}{\rm cm}}}\right)^2
  \left({L_{\rm UV} \over {10^{40}{\rm erg\, s^{-1}}}}\right)
  \left({D \over {0.1{\rm pc}}}\right)^{-2} \times \nonumber\\
  \times \min \left[{
      0.98\, \epsilon
      \left({m_{\rm p}\over{10^{-3}\,{\rm M_\odot}}}\right)^{-1}
      \left({r_{\rm p}\over{10^{10}\,{\rm cm}}}\right),\,
      (1-\epsilon{})}\right],\quad{}
\end{eqnarray}
where $m_{\rm prot}\simeq 1.67\times 10^{-24}\, {\rm g}$ is the proton mass. The bottom panel of Fig.~\ref{fig:fig2} (dotted line) shows the behavior of $\dot{M}$ as a function of the planet radius.

The average kinetic energy of the evaporating ions is of the order of $\sim \tilde{\phi}-\phi_0 = \epsilon \,{}[\phi_0/(1-\epsilon)]$, implying a velocity of the order of $v_{\rm g} \sim 30\,{\rm km\,{}s^{-1}}$ (for $\epsilon\sim{}0.3$). This velocity is generally of the same order of magnitude as the escape velocity $v_{\rm esc}=\sqrt{(2\,{}G\,{}m_{\rm p})/r_{\rm p}}$.

Thus, the density of the ionized gas that evaporates from the star is 
\begin{eqnarray}\label{eq:n0}
n_+=\frac{\dot{M}}{4\,{}\pi{}\,{}m_{\rm prot}\,{}r_{\rm p}^2\,{}v_{\rm g}}=10^{7}{\rm cm}^{-3}\left(\frac{\dot{M}}{6\times{}10^{10}{\rm g\,{}s^{-1}}}\right)\nonumber{}\\\,{}\left(\frac{30\,{}{\rm km s}^{-1}}{v_{\rm g}}\right)\,{}\left(\frac{10^{10}\,{}{\rm cm}}{r_{\rm p}}\right)^{2}.
\end{eqnarray}

The central panel of Fig.~\ref{fig:fig2} (dotted line) shows $n_+$ (obtained assuming $v_{\rm g}=v_{\rm esc}=\sqrt{(2\,{}G\,{}m_{\rm p})/r_{\rm p}}$), as a function of the planet radius.

If the velocity of the evaporating matter is close to the escape velocity from the planet, the density profile becomes $n(r)\sim n_+ (r/r_{\rm p})^{-3/2}$, and  the recombination rate can be estimated as 
\begin{eqnarray} 
R=\int_{r_{\rm p}}^{r_{\rm max}} 4\,{}\pi{}\,{} r^2 \alpha_{\rm B}\,{}n_+^2\,{}\left(\frac{r}{r_{\rm p}}\right)^{-3}{\rm d}r\nonumber{}\\=4\,{}\pi{}\,{}\alpha_{\rm B}\,{}n_+^2\,{} r_{\rm p}^3 \,{}\ln{r_{\rm max} \over r_{\rm p}}
\simeq 4\,{}\pi{}\,{}\alpha_{\rm B}\,{}n_+^2\,{} r_{\rm p}^3 \,{}\ln{\left({{n_+ \,{}m_{\rm prot}} \over \rho_{\rm h}}\right)^{2/3}}\nonumber{}\\
\sim{}3\times{}10^{33}{\rm s}^{-1}\left(\frac{n_+}{10^7{\rm cm}^{-3}}\right)^2\left(\frac{r_{\rm p}}{10^{10}{\rm cm}}\right)^3 {{\ln{\left[\frac{(n_+ \,{}m_{\rm prot})}{\rho_{\rm h}}\right]}}\over {9.7}},\nonumber{}\\
\label{eq:rec_slow}
\end{eqnarray}

where  $\alpha_{\rm B}\simeq 2.6\times10^{-13}\,{\rm cm^3\, s^{-1}}$ is the case B recombination coefficient for Hydrogen (at a temperature $\sim 10^4\,{\rm K}$), and we have chosen $r_{\rm max}$ as the radius where the density of the evaporated gas drops to the density of the hot medium: $m_{\rm prot}\,{}n(r_{\rm max}) \equiv \rho_{\rm h}$ (the normalization of the logarithmic term is appropriate for $n_+=10^7\,{\rm cm^{-3}}$, $\rho_{\rm h}=10^{-21}\,{\rm g\, cm^{-3}}$).  The corresponding emission measure (EM)  is EM$=\int{n_e^2\,{}{\rm d}V}\sim{}10^{45}$ cm$^{-3}$, for $r_{\rm p}=10^{10}$ cm and $n_+=10^{7}$ cm$^{-3}$.

 Equation (\ref{eq:rec_slow}) assumes that the evaporated gas is nearly completely ionized. This is a good approximation, because the recombination time scale
\begin{equation}
t_{\rm rec}(r) = {1\over{\alpha_{\rm B}n_+(r)}}
\simeq 3.8\times10^5\,{\rm s}\,
\left({{n_+}\over{10^7\,{\rm cm^{-3}}}}\right)^{-1}\,
\left({{r}\over{r_{\rm p}}}\right)^{3/2}
\end{equation}
is longer than the ionization time scale
\begin{equation}
t_{\rm ion} \sim {1\over{\sigma(\tilde{\phi}) L_{\rm UV}/(4\pi D^2 \tilde{\phi})}}
\simeq 1800\,{\rm s}\,
\left({{10^{40}\,{\rm erg\, s^{-1}}}\over{L_{\rm UV}}}\right)\,
\left({{D}\over{0.1\,{\rm pc}}}\right)^2,
\end{equation}
where we assumed that the the ionization cross section of neutral
Hydrogen is $\sigma(\tilde{\phi}) \simeq 6.3\times 10^{-18}\ {\rm cm^2}
(\tilde{\phi}/\phi_0)^{-3} \simeq 2.1\times 10^{-18}\, {\rm cm^2}$, which is appropriate for $\epsilon=0.3$, i.e. $\tilde{\phi}=\phi{}_0/0.7\simeq 19.4\,{\rm eV}$.

 Furthermore, in this case, we can ignore the possibility of shocks with the hot medium, because the relatively low values of $v_{\rm g}$ often lead to a Mach number $\mathcal{M}=v_{\rm g}/c_{\rm s}\lesssim 1$, where $c_{\rm s}$ is the sound speed  (slow-wind case). In the Appendix~\ref{sec:fast-wind}, we discuss the case in which $\mathcal{M}>>1$ (fast-wind case).  Photoevaporation in the fast-wind approximation leads to a recombination rate about one order of magnitude lower with respect to the slow-wind approximation, but shocks occurring in the fast-wind case enhance the recombination rate by a factor depending on $\mathcal{M}$ (see Appendix~\ref{sec:fast-wind}). 

In the following, we refer to CASE~1 (see Table~1) as the model where  $\dot{M}$, $n_+$ and $R$ have been calculated from equations~\ref{eq:mdot}, ~\ref{eq:n0} and \ref{eq:rec_slow}, respectively (i.e. $R$ has been calculated in the slow-wind approximation, with $v_{\rm g}=\sqrt{2\,{}G\,{}m_{\rm p}/r_{\rm p}}$ and  $L_{\rm UV}=10^{40}$ erg s$^{-1}$). Fig.~\ref{fig:fig2} (dotted line)  shows $\dot{M}$, $n_+$ and $R$  in CASE~1, as a function of the planet radius. 

\begin{deluxetable}{lllll}
\tabletypesize{\tiny}
\tablewidth{0pt}
\tablecaption{Summary of model properties.} 
\tablehead{
\colhead{Name} 
& \colhead{$\dot{M}$ (g s$^{-1}$)}
&  \colhead{$n_+$ (cm$^{-3}$)}
& \colhead{$R$  (s$^{-1}$)}
& \colhead{Tidal stripping}}
\startdata
CASE~1 & from eq.~\ref{eq:mdot} & from eq.~\ref{eq:n0} & from eq.~\ref{eq:rec_slow} & no \\
CASE~2 & from eq.~\ref{eq:mdotloeb} & from eq.~\ref{eq:n+loeb} & from eq.~\ref{eq:rec_slow} & no\\
CASE~3 & -- & -- & from eq.~\ref{eq:mdotstrip3}  & yes 
\enddata
\tablecomments{Columns 2, 3 and 4 specify how $\dot{M}$, $n_+$ and $R$ were calculated in each model. CASE~1 corresponds to a photoevaporating planet (eq.~\ref{eq:mdot}, Section~2) in the slow-wind approximation (eq.~\ref{eq:rec_slow}), with $m_{\rm p}=10^{-3}$ M$_\odot{}$, $v_{\rm g}=\sqrt{2\,{}G\,{}m_{\rm p}/r_{\rm p}}$ and $L_{\rm UV}=10^{40}$ erg s$^{-1}$. CASE~2 corresponds to a photoevaporating cloud (eq.~\ref{eq:mdotloeb}, Section~3) in the slow-wind approximation (eq.~\ref{eq:rec_slow}) with $m_{\rm p}=10^{-3}$ M$_\odot{}$, $v_{\rm g}=c_{\rm s}=10$ km s$^{-1}$ and $L_{\rm UV}=10^{40}$ erg s$^{-1}$. CASE~3 corresponds to a photoevaporating cloud undergoing tidal stripping ($r_{\rm p}\geq{}r_{\rm t}$), with  $v_{\rm g}=c_{\rm s}=10$ km s$^{-1}$ and $L_{\rm UV}=10^{40}$ erg s$^{-1}$. If the cloud is being stripped, we do not need to derive $\dot{M}$ and $n_+$ in order to calculate $R$, since $R=Q_{\rm tid}$  (see eq.~\ref{eq:mdotstrip3} and the discussion in the text).}
\end{deluxetable}

\section{Rogue protoplanetary embryos}\label{sec:PE}
One of the two main competing scenarios\footnote{In this paper, we neglect the competing core accretion model (i.e. the accretion of a gaseous atmosphere on a rocky core, e.g. \citealt{Wetherill80, Bodenheimer86, Lissauer93}), which will be considered in forthcoming studies.}
 for the formation of Jupiter-like planets and brown dwarfs predicts that they form through local gravitational instabilities in the outer parts of a protoplanetary disk (\citealt{Kuiper51,Boss97, Boss98a, Boss98b,Helled08}). The local instability might produce a protoplanetary embryo (PE): a coreless gas clump with density $\sim{}10^{-8}$ g cm$^{-3}$ (much lower than typical planetary and stellar densities), and large radius (corresponding to a Jeans length $\lambda{}_{\rm J}\approx{}$ few AU). Then, these gas clumps cool down, contracting to radii and densities typical of brown dwarfs or giant gaseous planets. Recent population synthesis studies \citep{ForganRice2013} indicate that PEs have a mass of a few to tens of Jupiter masses, and that a large fraction of PEs ($\approx{}40-90$ \%{}) end up forming brown dwarfs (with mass $\gtrsim{}0.02$ M$_\odot{}$).

One of the main predictions of this theory is that PEs can form only in the outer parts of protoplanetary disks ($\sim{}10-50$ AU distance from the star, \citealt{ForganRice2013}, or even $>100$ AU, \citealt{Boley09,Rafikov09}), where the self-gravity of gas is sufficiently strong with respect to stellar gravity and radiation pressure. Thus, a PE is an excellent candidate for being tidally captured by the SMBH, before it can contract to a Jupiter-like configuration. Furthermore, starless PEs might form directly from gravitational instabilities in a dense gaseous disk surrounding the SMBH (see previous section).
\subsection{Photoevaporation}\label{sec:PEphot}
For a radius of the PE $r_{\rm p}<r_{\rm t}\sim{}1.3\times{}10^{13}\,{}{\rm cm}\,{}(d/0.01\,{}{\rm pc})$ (see eq.~\ref{eq:rp}), the PE  will avoid tidal disruption during the capture by the SMBH. From eq.~\ref{eq:mdot}, we infer that the mass loss by evaporation for a PE is $\dot{M}\sim{}3.5\times{}10^{16}$ g s$^{-1}$  for $r_{\rm p}=5\times{}10^{12}$ cm (and $\epsilon=0.3$), corresponding to $n_+\sim{}2.9\times{}10^{8}$ cm$^{-3}$ for $v_{\rm g}=\sqrt{2\,{}G\,{}m_{\rm p}/r_{\rm p}}$. Thus, the recombination rate for an evaporating PE would be $R\sim{}3.1\times{}10^{44}$ s$^{-1}$ (using eq. \ref{eq:rec_slow}), corresponding to an EM $\sim{}10^{57}$ cm$^{-3}$. 
 However, this result neglects the optical depth encountered by the ionizing photons before reaching the PE surface,
\begin{eqnarray}
\tau = 
  \int_{r_{\rm p}}^{r_{\rm max}} \sigma(\tilde{\phi})\,
  n_+\, \left({r\over r_{\rm p}}\right)^{-3/2}
  {{t_{\rm ion}}\over{t_{\rm rec}(r)}}\, {\rm d}r \simeq \qquad{}\qquad{}\nonumber\\
  5\times 10^{-4} \left({n_+\over{10^7\,{\rm cm^{-3}}}}\right)^2
  \left({{r_{\rm p}}\over{10^{10}\,{\rm cm}}}\right)
  \left({{10^{40}\,{\rm erg\, s^{-1}}}}\over{L_{\rm UV}}\right)
  \left({D\over{0.1\,{\rm pc}}}\right)^2,
\label{eq:optdepth}
\end{eqnarray}
where we approximated the neutral fraction in the evaporating gas at
radius $r$ as $t_{\rm ion}/t_{\rm rec}(r)$ (which is correct as long
as $t_{\rm ion} \ll t_{\rm rec}(r)$).

 If we use the CASE~1 approximations at
$r_{\rm p} = 5 \times 10^{12}\, {\rm cm}$, we get $\tau \sim 200$,
implying that the estimate of $R$ is too high, since the ionizing
photons would be unable to photoevaporate the surface of the PE.
In general, our approximations break down when
$\tau \gtrsim 1$, i.e. when
$r_{\rm p} \gtrsim 1 {\rm AU} \,{}(n_+/10^7{\rm cm^{-3}})^{-2}\,{}
(L_{\rm UV}/10^{40}{\rm erg\,s^{-1}})\,{} (D/0.1{\rm pc})^{-2}$;
in particular, CASE~1 holds only if $r_{\rm p} \lesssim 2 \times
10^{11}\, {\rm cm}$.

 Therefore, it is more appropriate to derive the mass loss rate $\dot{M}$ with the assumption that the PE is a pure gas cloud (such as a protoplanetary disk). Following \citet{Murray-ClayLoeb13}, the mass loss rate of a gas cloud with radius $r_{\rm p}$, which is undergoing photoevaporation, can be expressed as
$\dot{M}\sim{}4\,{}\pi{}\,{}r_{\rm p}^2\,{}m_{\rm prot}\,{}n_+ \,{}c_{\rm s}$,
where 
the number density of the photoevaporated ions ($n_+$) can be calculated assuming balance between recombinations and photoionizations at the base of the wind,
\begin{equation}
n_+ \approx{}(L_{\rm UV}/(4\,{}\pi{}\,{}\tilde{\phi}\,{}D^2))^{1/2}\,{}(\alpha_{\rm B}\,{}r_{\rm p})^{-1/2},
\label{eq:n+loeb}
\end{equation}
where $\tilde{\phi}$ is the average energy of ionizing photons. Thus, the mass loss rate  is
\begin{eqnarray}\label{eq:mdotloeb}
\dot{M}\sim{}\left(\frac{4\,{}\pi{}}{\alpha_{\rm B}}\right)^{1/2}\,{}m_{\rm prot}\,{}c_{\rm s}\,{}r_{\rm p}^{3/2}\,{}\left(\frac{L_{\rm UV}}{\tilde{\phi}\,{}D^2}\right)^{1/2}\nonumber{}\\
\sim{}6.7\times{}10^{14}\,{}{\rm g}\,{}{\rm s}^{-1}\,{}\left(\frac{c_{\rm s}}{10\,{}{\rm km}\,{}{\rm s}^{-1}}\right)\,{}\left(\frac{r_{\rm p}}{10^{12}{\rm cm}}\right)^{3/2} \nonumber{}\\\,{}\left(\frac{L_{\rm UV}}{10^{40}\,{}{\rm erg}\,{}{\rm s}^{-1}}\right)^{1/2}\,{}\left(\frac{19.4\,{}{\rm ev}}{\tilde{\phi{}}}\right)^{1/2}\,{}\left(\frac{0.1\,{}{\rm pc}}{D}\right),
\end{eqnarray}
and the optical depth due to the evaporating gas becomes $\tau \sim 0.5$, independent of all the considered parameters, apart from $\tilde{\phi{}}$. This result shows that eq. (\ref{eq:n+loeb}) can be applied for a PE, since the absorption in the photoevaporative wind does not stop the photoevaporation itself.

 Furthermore, if we substitute eq. (\ref{eq:n+loeb}) into eq. (\ref{eq:rec_slow}), the recombination rate becomes
\begin{equation}
R \simeq {{4\pi r_{\rm p}^2 L_{\rm UV}} \over {4\pi D^2 \tilde{\phi}}}
\ln{\left({{r_{\rm max}} \over {r_{\rm p}}}\right)} =
\left[{4 \ln{\left({{r_{\rm max}} \over {r_{\rm p}}}\right)}}\right]
Q_{\rm p},
\label{eq:recPE}
\end{equation}
 where $Q_{\rm p}$ is the number of ionizing photons reaching the PE surface in the optically thin approximation (eq. \ref{eq:Qplanet}). 
We have $R>Q_{\rm p}$ because $r_{\rm max} \gg r_{\rm p}$, so that the number of available ionizing photons is much larger than $Q_{\rm p}$.


 The central and bottom panel of Fig.~\ref{fig:fig2} (solid line) show the behavior of $n_+$ and $\dot{M}$, as derived from equations~\ref{eq:n+loeb} and \ref{eq:mdotloeb}, respectively. Finally, the top panel of Fig.~\ref{fig:fig2} (solid line) shows the recombination rate $R$ obtained combining  eq.~\ref{eq:n+loeb}   with eq. \ref{eq:rec_slow} (or, equivalently, using eq. \ref{eq:recPE}). 
In the following, we define this model as CASE~2 (see Table~1).


\subsection{Tidal stripping enhancement}\label{subsec:strip}
The tidal radius $r_{\rm t}$ of a PE can become similar (or smaller) than its radius $r_{\rm p}$ for $d\lesssim 0.01$ pc (Fig.~\ref{fig:fig1}); in such case, mass loss by tidal stripping might become non-negligible at some point in the orbit. When $r_{\rm t}\le r_{\rm p}$, we can estimate the mass-loss rate by tidal stripping as (\citealt{Murray-ClayLoeb13})
\begin{eqnarray}\label{eq:mdotstrip}
\dot{M}_{\rm tid}\sim{}
4\pi r_{\rm p}^2\, \rho(r_{\rm p}) \left({m_{\rm p} \over {3M_{\rm BH}}}\right)^{1/3}\, v_\perp \simeq
r_{\rm p}^{-1}\, {m_{\rm p}^{4/3}\over {(3M_{\rm BH}})^{1/3}}\, v_\perp \simeq\nonumber\\
8.7\times{}10^{22}{\rm g\, s^{-1}}\, \left({r_{\rm p} \over {10^{12}{\rm cm}}}\right)^{-1}\, \left({m_{\rm p} \over {10^{-3}{\rm M_\odot}}}\right)^{4/3}\nonumber\\
\left({v_\perp \over {10^3{\rm km\, s^{-1}}}}\right)\, \left({M_{\rm BH} \over {4\times{}10^6{\rm M_\odot}}}\right)^{-1/3}
\end{eqnarray}
where we assume that the surface density of the PE is $\rho(r_{\rm p})\simeq m_{\rm p}/(4\pi r_{\rm p}^3)$ (appropriate in the case of an isothermal density profile), and $v_\perp$ is the radial component of the orbital velocity $v_{\rm p}$ (\citealt{Murray-ClayLoeb13}).

Which is the fate of the stripped material? How could we observe it? The tidally stripped material might undergo shocks with the high-temperature medium. The stagnation radius $r_{\rm s}$ where ram pressure is balanced between the bow shock of the stellar wind and the hot medium is \citep{Burkert12,Burkert13}
\begin{eqnarray}\label{eq:rshock}
r_{\rm s}=\left(\frac{\dot{M}\,{}v_{\rm g}}{4\,{}\pi{}\,{}\rho_{\rm h}\,{}v_{\rm p}^2}\right)^{1/2}= 2\times{}10^{16}\,{}{\rm cm}
\left(\frac{\dot{M}_{\rm tid}}{8.7\times{}10^{22}{\rm g\,{}s^{-1}}}\right)^{1/2}\nonumber{}\\\,
\left(\frac{v_{\rm g}}{10\,{}{\rm km\,{}s^{-1}}}\right)^{1/2}\,{}
\left(\frac{10^{-21}\,{}{\rm g\,{}cm^{-3}}}{\rho_{\rm h}}\right)^{1/2}\,{}
\left(\frac{1300\,{}{\rm km\,{}s}^{-1}}{v_{\rm p}}\right),\quad{}
\end{eqnarray} 
where $v_{\rm p}$ ($\sim 1300\,{\rm km\, s^{-1}}$) is the Kepler velocity of a planet orbiting the SMBH at 0.01 pc, and $\rho{}_{\rm h}=10^{-21}$ g cm$^{-3}$ is the density of the hot medium from X-ray measurements (\citealt{Yuan03}).

Thus, the maximum luminosity that can be emitted by the tidally stripped material in such shocks is
\begin{eqnarray}\label{eq:mdotstrip2}
L_{\rm max}=\pi{}\,{}r_{\rm s}^2\,{}\left(\frac{\rho_{\rm h}}{m_{\rm prot}}\,{}v_{\rm p}\right)\,{}\frac{1}{2}\,{}m_{\rm prot}\,{}v_{\rm p}^2=\frac{1}{8}\,{}\dot{M}_{\rm tid}\,{}v_{\rm g}\,{}v_{\rm p}\nonumber{}\\
\simeq{}1.4\times{}10^{36}{\rm erg\,{}s}^{-1}\,{}\left(\frac{\dot{M}_{\rm tid}}{8.7\times{}10^{22}\,{}{\rm g}\,{}{\rm s}^{-1}}\right)\nonumber{}\\\,{}\left(\frac{v_{\rm g}}{10\,{}{\rm km\,{}s}^{-1}}\right)\,{}\left(\frac{v_{\rm p}}{1300\,{}{\rm km\,{}s^{-1}}}\right),
\end{eqnarray}
where we used equations (\ref{eq:rshock}) and (\ref{eq:mdotstrip}) for $r_{\rm s}$ and $\dot{M}_{\rm tid}$, respectively. This luminosity is very high but can be achieved only if: (i) the stripped material reaches the stagnation radius $r_{\rm s}$ (but we estimate that the stripped gas needs $t\gtrsim 300$ yr to reach the stagnation radius, $r_{\rm s}\approx{}10^{16}{\rm cm}$, if it travels at $v_{\rm g}\sim{}10\,{}{\rm km\,{}s^{-1}}$); (ii) all of the kinetic energy is converted into (observable) luminosity. Thus, we expect that the luminosity due to these shocks is much lower than $L_{\rm max}$, and we will neglect it in the rest of this paper. 

On the other hand, the stripped material will be exposed to photoevaporation by the massive young stars (\citealt{Murray-ClayLoeb13}). The presence of the stripped material modifies the geometry of the PE and might strongly enhance photoevaporation (as discussed in \citealt{Murray-ClayLoeb13}). To correctly evaluate the new geometry is beyond the aims of the current paper, but we can account for this correction in the following way. The surface of the stripped material will take part in photoevaporation, and will produce an approximately spherical wind.

 The recombination rate in the wind will be of the same order of magnitude as the rate of ionizing photons that can be absorbed by the tidally stripped material, that is  
\begin{eqnarray}\label{eq:mdotstrip3}
Q_{\rm tid} = {{4 \pi{}\,{} r_{\rm str}^2\,{} L_{\rm UV}} \over {4 \pi D^2 \phi_0}},
\end{eqnarray}
where $r_{\rm str}$ is the maximum radius reached by the tidally stripped material. We evaluate $r_{\rm str}$ as
\begin{equation}\label{eq:eq_rstrip}
r_{\rm str} \sim r_{\rm p} + t_{\rm t} \,{}v_{\rm g},
\end{equation}
 i.e. as the radius of the PE, plus the distance travelled by the stripped material (moving at velocity $v_{\rm g}$) during the time $t_{\rm t}$ between the instant when the PE radius becomes equal to the tidal radius ($r_{\rm t}=r_{\rm p}$), and the time of the periapse passage.

\begin{figure}
\center{{
\epsfig{figure=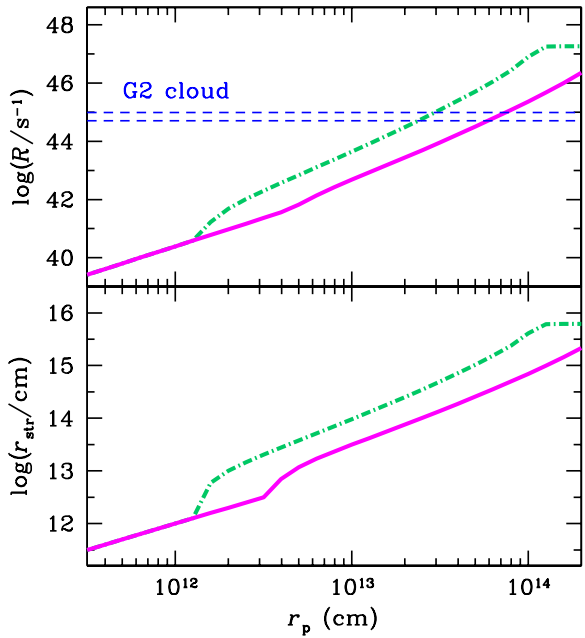,width=8.0cm} 
\caption{\label{fig:fig3}
 Bottom panel: radius $r_{\rm str}$ (eq.~\ref{eq:eq_rstrip}, 
assuming periapse distance $\sim{}200$ AU, eccentricity $e=0.976$ and velocity $v_{\rm g}=10$ km s$^{-1}$), as a function of PE radius. A value $r_{\rm str}=r_{\rm p}$ means that $r_{\rm p}$ is smaller than $r_{\rm t}$ for the entire orbit  (in this case, CASE~3 is the same as CASE~2). Top panel:  recombination rate $R$ in CASE~3 (Table~1), as a function of  PE radius. $R$ has been derived from eq.~\ref{eq:mdotstrip3}, assuming that $R=Q_{\rm tid}$. 
In both panels, magenta solid line: PE mass $m_{\rm p}=0.02$ M$_\odot$; green dot-dashed line: PE mass $m_{\rm p}=10^{-3}$ M$_\odot$.
 The two blue dashed lines in the top panel are the minimum and maximum value of the recombination rate measured for the G2 cloud in the Br$\gamma{}$ line since 2004 (\citealt{Pfuhl14}).
}}}
\end{figure}

Fig.~\ref{fig:fig3} shows the recombination rate\footnote{The calculations of
  \citet{Murray-ClayLoeb13} are somewhat similar to ours, even if they
  apply to a protoplanetary disk. However, \citet{Murray-ClayLoeb13}
  do not balance ionizations and recombinations. As illustrated by
  eq. (\ref{eq:recPE}), such balance is not required in the case of
  planets or PEs, because $r_{\rm max} \gg r_{\rm p}$. However, when
  the tidal stripping enhancement is important, the ratio $r_{\rm
    max}/r_{\rm str}$ should be of the order of unity, and $R \sim
  Q_{\rm tid}$. For example, in the case of
  disks with radius $\gtrsim 10\,{\rm AU}$ producing photoevaporative
  winds with $n\sim10^{7}\,{\rm cm^{-3}}$, the equations used by
  \citet{Murray-ClayLoeb13} imply a recombination rate much larger
  than the ionization rate, and the material would become neutral in
  $\sim5$ days.}  $R$ (assumed to be
equal to $Q_{\rm tid}$) and the radius $r_{\rm str}$, as derived from equations \ref{eq:mdotstrip3} and \ref{eq:eq_rstrip},
respectively. 
 In Fig.~\ref{fig:fig3}, we assume $v_{\rm g}=10$ km
s$^{-1}$, periapse distance $=200$ AU and eccentricity $e=0.976$,
i.e. the same periapse and eccentricity as the G2 object.  
 If $r_{\rm p}>r_{\rm t}$ along the
entire orbit, we assume $t_{\rm t}=0.5\,{}T_{\rm orb}$ (where $T_{\rm
  orb}$ is the orbital period). In principle, this kind of calculation
can be applied also to planets, but for the orbit of the G2 cloud
$r_{\rm t}$ is always larger than $r_{\rm p}$ if $r_{\rm
  p}\lesssim{}1.5\times{}10^{12}$ cm (for $m_{\rm p}=10^{-3}$
M$_\odot$). In the following, we refer to the model presented in
Fig.~\ref{fig:fig3} as CASE~3 (see Table~1).

\section{Discussion}\label{sec:discussion}
\subsection{Luminosity of rogue planets in the GC}
Do we have any chances of detecting rogue planets or PEs in the GC with current or forthcoming facilities?
The $K$ (2.1 $\mu{}m$) and $L'$ (3.8 $\mu{}m$) magnitudes of a rogue planet at the distance of the GC ($\sim{}8$ kpc) are $m_{\rm K}\sim{}32$ and  $m_{\rm L'}\sim{}26$ for a mass $m_{\rm p}\sim{}10^{-3}$ M$_\odot$ (\citealt{Allard01, Allard07}), respectively. Therefore, such rogue planet would be invisible for current and forthcoming facilities (30-m class telescopes are expected to observe stars down to $m_{\rm K}\sim{}24$). On the other hand, some processes might take place that enhance the chances of observing a planet, such as tidal disruption, atmosphere evaporation and bow shocks. In the previous section, we analyzed these processes for both planets and PEs.

\begin{figure}
\center{{
\epsfig{figure=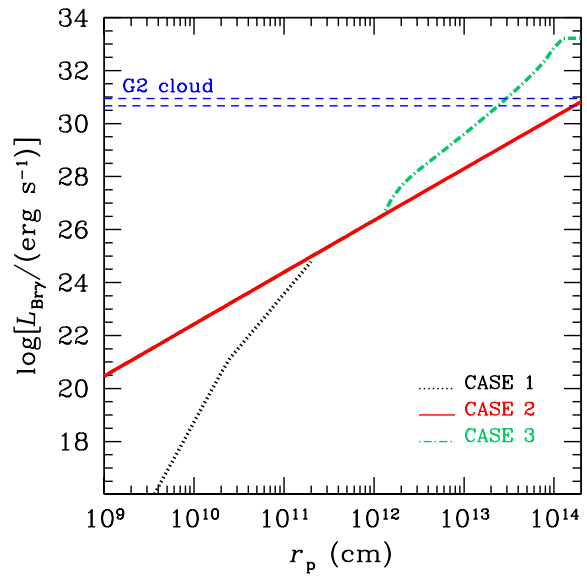,width=8.0cm} 
\caption{\label{fig:fig4}
 Br$\gamma{}$ luminosity ($L_{{\rm Br}\gamma{}}$) as a function of the planet (or PE) radius. The lines and colors are the same as used in Figs.~\ref{fig:fig2} and ~\ref{fig:fig3}. In particular, dotted black line: CASE~1; 
solid red line: CASE~2; green dash-dotted line: CASE~3 with $m_{\rm p}=10^{-3}$ M$_\odot$. The two blue dashed lines are the minimum and maximum value of $L_{{\rm Br}\gamma{}}$ for the G2 cloud observed since 2004 (\citealt{Pfuhl14}).
}}}
\end{figure}

Fig.~\ref{fig:fig4} shows the luminosity of the Br$\gamma{}$ line (2.166 $\mu{}$m) derived from $L_{{\rm Br}\gamma{}}=2.35\times{}10^{-27}\,{\rm erg\, s^{-1}}\, R/\alpha{}_{\rm B}$,  where the recombination rate $R$ was calculated in the previous section. 
The most optimistic prediction for a Jupiter-like planet ($m_{\rm p}=10^{-3}$ M$_\odot$ and $r_{\rm p}=10^{10}$ cm) is a Br$\gamma$ luminosity $L_{{\rm Br}\gamma{}}\approx{}5\times{}10^{18}$ erg s$^{-1}$, if the planet atmosphere is undergoing photoevaporation. At the distance of the GC, this corresponds to a flux $\lesssim 10^{-27}\,{\rm erg\, s^{-1}\, cm^{-2}}$, which is far below the sensitivity of existing and forthcoming instruments (a 24-hr integration with VLT SINFONI can theoretically detect a Br$\gamma$ flux $\sim5\times10^{-18}\,{\rm erg\, s^{-1}\, cm^{-2}}$ with a S/N of $\sim10$). 

Thus, the only chance of detecting a rogue planet in the GC is that it is disrupted and accretes onto the SMBH. If a portion of the planet mass is accreted by the SMBH, this might lead to a flare with bolometric luminosity $\le{}2\times{}10^{41}$ erg s$^{-1}$ (from eq.~45 of \citealt{Zubovas12}). 
This event is very unlikely, because the tidal disruption occurs only if the  periapse distance from the SMBH is $\lesssim{}1.6\,{}{\rm AU}\,{}(r_{\rm p}/10^{10}{\rm cm})$. 
 We expect a planet disruption rate $R_{\rm dis}\sim{}10^{-5}f_{\rm p}$ yr$^{-1}$ (where $f_{\rm p}$ is the number of planets per star in the GC, and $10^{-5}$ yr$^{-1}$ is the tidal disruption rate of stars estimated by \citealt{Alexander05}). 
This process is very rare, but might be relevant for explaining flares in other galactic nuclei.

\subsection{Luminosity of rogue PEs in the GC}
Theoretical models (e.g. \citealt{Wuchterl99, Boss05, Helled06, Helled10})  suggest that a 0.001 M$_\odot$ PE  does not collapse immediately (because it is optically thick) and  needs to cool for $\lesssim{}1$ Myr at a bolometric luminosity $\sim{}10^{-6}$ L$_\odot{} \sim 10^{27}\, {\rm erg\, s^{-1}}$. At the distance of the GC, this luminosity  corresponds to $m_{\rm L'}\gtrsim 25$, which is far below the limits of current and forthcoming observational facilities.

The photoevaporation of the `atmosphere' of a PE (with a radius of $r_{\rm p}=5\times{}10^{12}$ cm) could lead to a  Br$\gamma$ luminosity  of $\sim{}5\times{}10^{27}$erg s$^{-1}$, considering CASE~2 estimates for the mass loss by photoevaporation, assuming $L_{\rm UV}=10^{40}$ erg s$^{-1}$. 
 If the PE is undergoing tidal stripping, this luminosity might be boosted by more than 2 orders of magnitude. 
A $L_{{\rm Br}\gamma{}}\approx{}10^{30}$ erg s$^{-1}$ can be observed with current 8m telescopes, and is not far from the actual luminosity of the G2 cloud ($\sim{}6\times{}10^{30}$ erg s$^{-1}$).

Tidal disruption of a PE occurs if the distance of the PE from the SMBH is $d\sim{}750\,{}{\rm AU}\,{}(r_{\rm p}/5\times{}10^{12}{\rm cm})$, suggesting that the cross section for PE disruption is a factor of $\approx{}10^5$ larger than the cross section for planet disruption (neglecting gravitational focusing). This leads to a disruption rate of $\sim{}f_{\rm PE}$ yr$^{-1}$, where $f_{\rm PE}$ is the number of PEs per each star. 

How many PEs can form in the GC and for how long do they survive? This question contains a number of major uncertainties and we can just suggest a few hints. PEs are quite elusive objects and are expected to be relatively short-lived before they contract to Jupiter-like size ($\sim{}10^{3-6}$ yr, \citealt{Wuchterl99,Helled08,ForganRice2013}), but in the GC they could be efficiently separated from their parent star, and the strong UV flux might substantially slow down their cooling (and collapse). Furthermore, we could reverse our question, and use the non-detection of $L_{{\rm Br}\gamma{}}\approx{}10^{31}$ erg s$^{-1}$ objects to constrain the frequency of PEs in the GC. 

\subsection{Rogue PEs and the G2 cloud}
The G2 cloud is the only observed object (so far) that shares similar properties with a photoevaporating and partially stripped PE.  
G2 probably originated from the clockwise disk in the GC, and has a very high eccentricity, $e\sim{}0.976\pm{}0.007$ (\citealt{Pfuhl14}). Both these constraints on the orbit are fairly consistent with the hypothesis of tidal capture of a PE (initially formed in the clockwise disk) by the SMBH, or of a dynamical ejection of the PE from the clockwise disk. It is now clear that G2 survived its periapse passage at a distance of $\sim{}200$ AU from the SMBH (\citealt{Witzel14}), but some of its material was tidally stripped. A PE with radius $\sim{}10^{12}$ cm is tidally stripped if the periapse is $\sim{}160$ AU: such PE would not be completely tidally disrupted at a distance of $\sim{}200$ AU from the SMBH, but it would suffer some tidal stripping.

Fig.~\ref{fig:fig4} shows that the  Br$\gamma{}$ luminosity of a PE (as derived from photoevaporation and partial tidal stripping) matches the one observed for G2, if the PE has $m_{\rm p}\approx{}10^{-3}$ M$_{\odot}$ and  $r_{\rm p}\approx{}2\times{}10^{13}$ cm. If the main trigger of the Br$\gamma{}$ emission is photoevaporation by the intense UV background in the GC, we expect the Br$\gamma{}$ luminosity to remain roughly constant during the PE orbit. This is fairly consistent with the fact that the Br$\gamma{}$ luminosity of G2 remained nearly constant over 10 yr (\citealt{Pfuhl14}). 

G2 was detected also in  the $L'$ band ($m_{\rm L'}\sim{}14$, \citealt{Pfuhl14}). \citet{Gillessen12} suggest that the continuum $L'$ emission comes from small ($\sim{}20$ nm), transiently heated dust grains with a total warm dust mass of $\sim{}2\times{}10^{23}$ g. The grains might be warmed up by an inner source (e.g. a low-mass star, \citealt{Burkert13,Ballone13, Witzel14}),  or by some external mechanism (UV heating, shocks, etc.). 
In the PE scenario, the minimum PE radius necessary to  reach $L_{\rm L'}\sim{}2.1\times{}10^{33}{\rm erg\,{}s^{-1}}$ (\citealt{Witzel14}) is
\begin{eqnarray}\label{eq:Lprime}
 r_{\rm min}\sim{}5.5\times{}10^{12}{\rm cm}\left(\frac{L_{\rm L'}}{2.1\times{}10^{33}{\rm erg\,{}s^{-1}}}\right)^{1/2}\,{}\left(\frac{T_{\rm dust}}{560\,{}{\rm K}}\right)^{-2},
\end{eqnarray}
where $T_{\rm dust}\sim{}560$ K is the estimated dust temperature (from $L'-M'\sim{}0.3$, \citealt{Gillessen12}). We note that $r_{\rm min}$ strongly depend on $T_{\rm dust}$ and that the estimate of $T_{\rm dust}$ is very uncertain\footnote{It should be noted that neither the radiation absorbed by the PE ($\sim 10^{31}\,{\rm erg\, s^{-1}}\, [L_{\rm UV}/(10^{40}\,{\rm erg\, s^{-1}})]\, [D/(0.1\,{\rm pc})]^{-2} [r_{\rm p}/(10^{13}\,{\rm cm})]^2$), nor the energy released in shocks around it ($\le 2.5\times10^{30}\, {\rm erg\, s^{-1}}\, [\rho_{\rm h}/(10^{-21}\,{\rm g\, cm^{-3}})]\, [r_{\rm p}/(10^{13}\,{\rm cm})]^2\, [v_{\rm p}/(1000\,{\rm km\, s^{-1}})]^3$) can balance with $L_{\rm L'}$, unless $r_{\rm p}\gtrsim 5\times10^{13}\, {\rm cm}$.}.

While the scenario of a wind-enshrouded low-mass star would naturally explain the continuum $L'$ emission (\citealt{Witzel14}), we cannot reject the hypothesis that a PE, embedded in the hot dense medium of the GC, might host sufficient warm dust to power the observed $L'$ emission. Thus, a rogue PE might be a viable scenario to explain G2 and other G2-like objects (e.g. \citealt{Pfuhl14} suggest that the object G1 is related to G2). 

\subsection{Discussion of uncertainties}
\begin{figure}
\center{{
\epsfig{figure=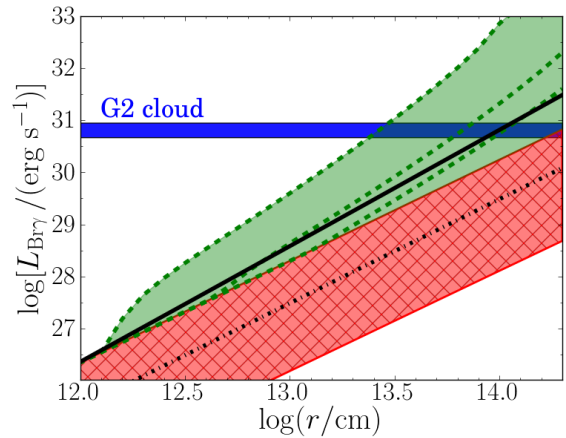,width=8.0cm} 
}}
\caption{\label{fig:fig5}
   Uncertainties on the Br-$\gamma{}$ luminosity of a PE. Red filled cross-hatched area: CASE~2 by varying $L_{\rm UV}$ from $10^{38}$ erg s$^{-1}$ (lower bound) to $10^{40}$ erg s$^{-1}$ (upper bound). 
Dot-dashed black line: same as CASE~2 with $L_{\rm UV}=10^{40}$ erg s$^{-1}$ but in the fast-wind approximation (with $v_{\rm g}=10\,{}c_{\rm s}$, see Appendix~\ref{sec:fast-wind}, eq.~\ref{eq:rec}).  Solid black line: Br-$\gamma{}$ luminosity from shocks in the fast-wind approximation (with $v_{\rm g}=10\,{}c_{\rm s}$, see Appendix~\ref{sec:fast-wind}, eq.~\ref{eq:shock}). 
Green filled area: CASE~3 with $L_{\rm UV}=10^{40}$ erg s$^{-1}$,  by varying the PE mass from $10^{-3}$ M$_\odot{}$ (upper bound) to $\sim{}0.2$ M$_\odot{}$ (lower bound). Dashed green lines: $m_{\rm p}=10^{-3}$ M$_\odot{}$ (top), $0.02$ M$_\odot{}$ (intermediate), $0.1$ M$_\odot{}$ (bottom).  Horizontal blue filled area: the observed Br-$\gamma{}$ luminosity of the G2 cloud (\citealt{Pfuhl14}).}
\end{figure}
 The analytic model discussed in this paper relies on the possibility that planets form in the GC. The GC might be a hostile environment for planet formation, given the high UV background, the high temperature and the strong tidal field by the SMBH. Discussing the chances that planets form in the GC is beyond the aims of this paper. Here, we just mention that planetesimals and asteroids were recently investigated as possible sources of SMBH flares \citep{Cadez08, Kostic09, Zubovas12, Hamers14}, and there are even some observational hints for the existence of protoplanetary disks in the innermost parsec \citep{Yusef15}. Further observational evidence is necessary, to confirm that planets and planetary objects form in the GC.

Planets cannot  be directly observed with current and forthcoming facilities. Only photoevaporating PEs are sufficiently bright to be detected. PEs are theoretically predicted objects but have not yet been observed. They can form only if gravitational instabilities in a protoplanetary disk lead to fragmentation, and to the formation of self-gravitating clumps. 
Recent work has shown that gravitational instabilities can lead to the formation of PEs only in the outermost regions of a protoplanetary disk ($>100$ AU, \citealt{Boley09}), where self-gravity is stronger. This enhances the probability that PEs become unbound with respect to their initial system, but might imply that their formation is endangered by the strong UV field in the GC. Furthermore, even if  hydrodynamical simulations show that clumps can form, they cannot predict self-consistently whether these clumps survive further evolution. 

Recent estimates suggest that PEs remain in the pre-collapse phase (the one considered in this paper) for $\sim{}10^5$ yr and for $\sim{}10^{4}$ yr if their mass is $\sim{}10^{-3}$ and $10^{-2}$ M$_\odot$, respectively \citep{Helled06,Helled10}. From an observational perspective, while the detection of massive giant planets ($\gtrsim{}10^{-3}$ M$_\odot{}$) at distance $>>10$ AU from the central star might favor the gravitational instability scenario (\citealt{Marois08}; see Figure~3 of \citealt{Pepe14}), there is no strong evidence supporting the existence of PEs. Since PEs are such elusive objects, it is quite difficult to quantify the uncertainties in our model. 

Besides the debate on the very existence of PEs, Fig.~\ref{fig:fig5} shows the impact  of the main sources of uncertainties on the predicted Br-$\gamma{}$ luminosity of a photoevaporating PE. In particular, we focus on (1) the adopted approximation for the wind (fast or slow-wind approximation),  (2) the flux of the UV background (from $10^{38}$ erg s$^{-1}$ to $10^{40}$ erg s$^{-1}$ in the innermost 0.1 pc), (3) the mass of the PE. 

Fig.~\ref{fig:fig5} shows that the Br-$\gamma{}$ luminosity of a photoevaporating PE is uncertain by several orders of magnitude. We stress that the recombination rate (and thus the Br-$\gamma{}$ luminosity) does not depend on the mass of the PE in CASE~2. The mass of the PE is important only if we assume that the PE can be tidally stripped by the SMBH (CASE~3), because $r_{\rm t}$ depends on the PE mass. 
In the fast-wind approximation (i.e. the gas is ejected with $v_{\rm g}>>v_{\rm esc}$, as discussed in detail in Appendix~\ref{sec:fast-wind}), the Br-$\gamma{}$ luminosity due to photoevaporation (from eq.~\ref{eq:rec}) is lower by about one order of magnitude with respect to the slow-wind approximation, but the gas may undergo shocks with the hot medium, and this enhances the Br-$\gamma{}$ luminosity to a value (from eq.~\ref{eq:shock}) comparable with (or even higher than) the slow-wind approximation.


\section{Summary}\label{sec:summary}
 In this paper, we investigated the possible observational signatures of planets and planetary embryos (PEs) in the GC. If planets and PEs form in the central parsec, several mechanisms can separate them from their parent star (e.g. tidal shear by the SMBH or planet-planet scatterings), and bring them onto a very eccentric orbit around the SMBH. It is even possible that starless PEs and planets form directly from gravitational instabilities in a dense gas disk around the SMBH (such as the one that might have given birth to the clockwise disk of young massive stars).

We have shown that both planets and PEs suffer from photoevaporation (Fig.~\ref{fig:fig2}) due to the intense UV background. The emission measure associated with such process is relatively low for planets (EM$\sim{}10^{45}$ cm$^{-3}$) and much higher, although very uncertain, for PEs (EM$\approx{}10^{50-56}$ cm$^{-3}$). In the case of PEs, tidal stripping can enhance the effect of photoevaporation, leading to an even higher EM (up to EM$\approx{}10^{60}$ cm$^{-3}$, Fig.~\ref{fig:fig3}). This means that a photoevaporating PE with radius  $\sim{}5\times{}10^{13}$ cm might reach a Br-$\gamma{}$ luminosity $L_{\rm Br\gamma{}}\approx{}5\times{}10^{29}$ erg s$^{-1}$ if it is not tidally stripped, and  $L_{\rm Br\gamma{}}\approx{}5\times{}10^{31}$ erg s$^{-1}$ if it is partially tidally stripped (Fig.~\ref{fig:fig4}). This value, while uncertain, is remarkably similar to the observed Br-$\gamma{}$ luminosity of the G2 dusty object. Furthermore, a PE with radius $\gtrsim{}5\times{}10^{13}$ cm can emit the same $L_{\rm L'}$ luminosity as G2, if it contains the sufficient amount of dust at temperature $\approx{}600$ K. In our model, the $L'$ luminosity is emitted from a smaller area than the Br-$\gamma{}$ line, since the former is due to dust inside the PE, while the latter is produced by the photoevaporative wind. This can account for the fact that the observed $L'$ emission is more compact than the emission in Br-$\gamma{}$ \citep{Witzel14}. If  G2  is a PE with radius $\sim{}5\times{}10^{13}$ cm, we expect its lifetime to be $t_{\rm life}\approx{}10^5$ yr $(m_{\rm p}/10^{-3} {\rm M}_\odot{})$ $(5\times{}10^{17}{\rm g\,{}s}^{-1}/\dot{M})$, but tidal stripping can reduce $t_{\rm life}$ significantly (down to $\approx{}100$ yr).


Our results are affected by several uncertainties. First, PEs are theoretically predicted objects, but elusive to observe. Their properties and survival time are  uncertain. Furthermore, the luminosity of photoevaporating PEs strongly depends on several quantities (e.g. PE mass, UV background luminosity, wind speed), as shown in Fig.~\ref{fig:fig5}. In a follow-up work we will investigate the hydrodynamical evolution of PE models embedded in a UV background. Furthermore, the frequency of PEs in the GC, and the probability that they are captured by the SMBH deserves further study. Our preliminary results open a new exciting window on GC's environment.

\section*{Acknowledgments}
We thank Mariangela Bonavita and Alessia Gualandris for useful discussions. MM acknowledges financial support from the Italian Ministry of Education, University and Research (MIUR) through grant FIRB 2012 RBFR12PM1F, and from INAF through grants PRIN-2011-1 and PRIN-2014-14. ER acknowledges financial support from Progetto di Ateneo 2012, University of Padova, ID: CPDA125588/12.

\appendix
\section{The fast-wind approximation}\label{sec:fast-wind}
 In the main text, we discussed the case in which $v_{\rm g}\lesssim{}v_{\rm esc}$. While this is the more likely scenario, in this appendix we also consider the case in which the initial velocity $v_{\rm g}>>v_{\rm esc}$. If $v_{\rm g}>>v_{\rm esc}$, we can assume that $v_{\rm g}$ remains approximately constant, so that the gas density scales as $n(r)\sim{}n_+\,{}(r/r_{\rm p})^{-2}$. Thus, we can estimate the recombination rate as
as
\begin{eqnarray}\label{eq:rec}
R=\int_{r_{\rm p}}^r 4\,{}\pi{}\,{} r_{\rm max}^2 \alpha_{\rm B}\,{}n_+^2\,{}\left(\frac{r}{r_{\rm p}}\right)^{-4}{\rm d}r\nonumber{}\\
=4\,{}\pi{}\alpha_{\rm B}\,{}n_+^2\,{} r_{\rm p}^3 \,{}\left(1 - \frac{r_{\rm p}}{r_{\rm max}}\right)\nonumber{}\\\sim{}3.3\times{}10^{32}{\rm s}^{-1}\left(\frac{n_+}{10^7{\rm cm}^{-3}}\right)^2\left(\frac{r_{\rm p}}{10^{10}{\rm cm}}\right)^3,
\end{eqnarray}
where $\alpha_{\rm B}\sim{}2.6\times{}10^{-13}$ cm$^{3}$ s$^{-1}$ is the Case B radiative recombination coefficient for H (at a temperature of $\sim10^4$ K), and $r_{\rm max}$ (typically $\gg r_{\rm p}$) is the outer limit of the wind-dominated density profile (either the radius where the interaction between the wind and the high-temperature medium produces a shock - see below -, or the radius where the wind density drops below that of the surrounding medium).
A recombination rate $R\sim{} 1.2\times10^{33}{\rm s}^{-1}$ corresponds to an emission measure EM=$\int{}n^2_e{\rm d}V\sim{} 3\times10^{46}$ cm$^{-3}$. 

The wind that evaporates from the planet will also undergo a shock with the high-temperature medium in the GC. Equation~\ref{eq:rshock} in Section~\ref{subsec:strip} provides  the stagnation radius $r_{\rm s}$ where ram pressure is balanced between the bow shock of the stellar wind and the hot medium  \citep{Burkert12,Burkert13}. 

We estimate the combined effect of the shock and the photoevaporation with a simplified version of the results of \cite{Dyson75}: we assume that the results along the direction of motion 
are approximately valid for all directions, obtaining a recombination rate
\begin{eqnarray}\label{eq:shock}
\tilde{R}\simeq{}4\,{}\pi{}\,{}\alpha_{\rm B}\,{}n_+^2\,{} r_{\rm p}^3\,{}\left(1+\frac{r_{\rm p}}{r_{\rm s}}\mathcal{M}^2\right)\nonumber{}\\
\sim{}2.5\times{}10^{33}{\rm s}^{-1}\left(\frac{n_+}{10^7{\rm cm}^{-3}}\right)^2\left(\frac{r_{\rm p}}{10^{10}{\rm cm}}\right)^3.
\end{eqnarray}
Equation \ref{eq:shock} holds only if  $\mathcal{M}^2>>1$, where $\mathcal{M}\equiv{}v_{\rm g}/c_{\rm s}$ is the Mach number, and $c_{\rm s}$ is the sound speed (the normalization used in eq. \ref{eq:shock} adopts $\mathcal{M}=5$, corresponding to $v_{\rm g}=50$ km s$^{-1}$, and $c_{\rm s}=10$ km s$^{-1}$). Thus, the contribution of shocks increases the recombination rate by a factor of $\approx{}8$ for $\mathcal{M}=5$. In Fig.~\ref{fig:fig5}, we compare the Br-$\gamma{}$ luminosity of a PE in the slow-wind approximation and in the fast-wind approximation. 


{}
\end{document}